\documentclass[aps,twocolumn,pra,superscriptaddress,showpacs,tightenlines,pdflatex,longbibliography]{revtex4-1}
\usepackage{amssymb}
\usepackage{amsmath}
\usepackage{dcolumn}
\usepackage{graphicx}
\usepackage{mathrsfs}
\usepackage{appendix}
\usepackage{graphicx}
\usepackage{booktabs}
\usepackage{color}
\usepackage{url}
\usepackage[colorlinks]{hyperref}

\hypersetup{    plainpages=true,
    breaklinks=true,    hypertexnames=false,    pageanchor=true,
    colorlinks=true,
    linkcolor={blue},
    citecolor={blue},
    urlcolor={blue},
    anchorcolor={black}
}

\begin{document}

\title{Observing pure effects of counter-rotating terms without ultrastrong
coupling:\\ A single photon can simultaneously excite two qubits}
\author{Xin Wang}
\affiliation{Institute of Quantum Optics and Quantum Information,
School of Science, Xi'an Jiaotong University, Xi'an 710049, China}
\affiliation{CEMS, RIKEN, Wako-shi, Saitama 351-0198, Japan}

\author{Adam Miranowicz}
\email{miran@amu.edu.pl}
\affiliation{CEMS, RIKEN, Wako-shi, Saitama 351-0198, Japan}
\affiliation{Faculty of Physics, Adam Mickiewicz University,
61-614 Pozna\'n, Poland}

\author{Hong-Rong Li}
\email{hrli@mail.xjtu.edu.cn}
\affiliation{Institute of Quantum Optics and Quantum Information,
School of Science, Xi'an Jiaotong University, Xi'an 710049, China}

\author{Franco Nori}
\affiliation{CEMS, RIKEN, Wako-shi, Saitama 351-0198, Japan}
\affiliation{Physics Department, The University of Michigan, Ann
Arbor, Michigan 48109-1040, USA}
\date{\today}

\begin{abstract}
The coherent process that a single photon simultaneously excites
two qubits has recently been theoretically predicted by
[\href{https://link.aps.org/doi/10.1103/PhysRevLett.117.043601}{Phys.
Rev. Lett. 117, 043601 (2016)}]. We propose a different approach
to observe a similar dynamical process based on a superconducting
quantum circuit, where two coupled flux qubits longitudinally
interact with the same resonator. We show that this simultaneous
excitation of two qubits (assuming that the sum of their
transition frequencies is close to the cavity frequency) is
related to the counter-rotating terms in the dipole-dipole
coupling between two qubits, and the standard rotating-wave
approximation is not valid here. By numerically simulating the
adiabatic Landau-Zener transition and Rabi-oscillation effects, we
clearly verify that the energy of a single photon can excite two
qubits via higher-order transitions induced by the longitudinal
couplings and the counter-rotating terms. Compared with previous
studies, the coherent dynamics in our system only involves one
intermediate state and, thus, exhibits a much faster rate. We also
find transition paths which can interfere. Finally, by discussing
how to control the two longitudinal-coupling strengths, we find a
method to observe both constructive and destructive interference
phenomena in our system.
\end{abstract}

 \maketitle

\affiliation{Institute of Quantum Optics and Quantum Information,
School of Science, Xi'an Jiaotong University, Xi'an 710049, China} %
\affiliation{CEMS, RIKEN, Wako-shi, Saitama 351-0198, Japan}

\affiliation{CEMS, RIKEN, Wako-shi, Saitama 351-0198, Japan}
\affiliation{Faculty of Physics, Adam Mickiewicz University,
61-614 Pozna\'n, Poland}

\affiliation{Institute of Quantum Optics and Quantum Information,
School of Science, Xi'an Jiaotong University, Xi'an 710049, China}

\affiliation{CEMS, RIKEN, Wako-shi, Saitama 351-0198, Japan}
\affiliation{Physics Department, The University of Michigan, Ann
Arbor, Michigan 48109-1040, USA}

\section{Introduction}
The light-matter interaction between a quantized electromagnetic
field and two-level atoms has been the central topic of quantum
optics for half a century, and has developed into the standard
cavity quantum electrodynamics (QED) theory. In a QED system, if
the dipole-field or dipole-dipole coupling strengths ($\lambda$)
are weak compared with the cavity or atomic transition frequencies
($\omega_{c}$ and $\omega_{q}$, respectively), we often routinely
adopt the rotating-wave approximation (RWA). Under the RWA, one
can neglect the excitation-number-nonconserving
terms~\cite{Bloch40,Jaynes63,Bruce93,Irish07}, which, compared
with the resonant terms, are usually only rapidly oscillating
virtual processes and negligibly contribute to the dynamical
evolution of such a system~\cite{Scully1997}.

In fact, the RWA works well even in the strong-coupling regime.
Only in the ultrastrong and deep-strong coupling regimes [where
$\lambda>0.1\times\min{(\omega_{c}, \omega_{q})}$ and
$\lambda>\min{(\omega_{c}, \omega_{q})}$,
respectively]~\cite{Anappara09, FornDiaz2010, Niemczyk2010,
FornDiaz2010, Geiser12,Scalari2012,Baust2016, FornDiaz2017,
Yoshihara2017, Chen17}, the counter-rotating terms have apparent
effects in a QED system~\cite{Braak11,Gu2017}. The
excitation-number-nonconserving terms in a QED system can lead to
many interesting quantum
effects~\cite{Niemczyk2010,Ashhab10,Cao10,Cao2011a,Cao2012,Casanova10,Garziano15,Wang16},
such as three-photon resonances~\cite{Ma15}, the modification of
the standard input-output relation~\cite{Ridolfo12, Ridolfo13},
quantum phase transitions~\cite{Hwang2015}, frequency
conversion~\cite{Kockum2017b}, or the deterioration of photon
blockade effects~\cite{Hwang16,Alexandre16}. However, all of these
phenomena are the combined and mixed effects of both
counter-rotating and resonant terms. Here we address, in
particular, the following questions: How can we observe some pure
effects of counter-rotating wave terms in a QED system, i.e.,
without being disturbed by the resonant terms? Moreover, is it
really always reasonable to apply the RWA in dipole-field or
dipole-dipole coupling systems, which are far way from the
ultrastrong-coupling regime?

Other interesting quantum processes are multi-excitation and
emission in a QED system. The process that a two-level atom
(molecule) absorbs two or more photons simultaneously has been
widely discussed in many quantum
platforms~\cite{Denk90,Peter00,Garziano15,Wang16,Chen17}. However,
the inverse process (of a single photon splitting to excite two
and more atoms) is rarely
studied~\cite{Garziano2016,Kockum2017a,Stassi2017}.

Recently, Garziano \emph{et al.}~\cite{Garziano2016} predicted
that \emph{one photon can simultaneously excite two or more
qubits}. In their theoretical proposal, two superconducting qubits
are coupled to a resonator with both longitudinal and transverse
forms in the ultrastrong-coupling regime. A similar process was
predicted via the photon-mediated Raman interactions between two
three-level atoms (qutrits) in the strong-coupling
regime~\cite{Zhao2017}. Note that both dynamics (with
qubits~\cite{Garziano2016} and qutrits~\cite{Zhao2017}) were
composed of three virtual processes, which do not conserve the
number of excitations. Also the effective transition between atoms
and a single photon is of a relatively slow rate.

In this paper, we propose a superconducting system composed of a
transmission-line resonator longitudinally coupled with two flux
qubits. The two qubits couple to each other via an
antiferromagnetic dipole-dipole interaction. We show that, when
the sum of two qubits transition frequencies is approximately
equal to the resonator frequency, the counter-rotating terms in
the dipole-dipole interaction cannot be dropped even when the
system has not entered into the ultrastrong-coupling regime. The
RWA is not valid here; on the contrary, the resonant terms can be
approximately neglected in our model. Due to the counter-rotating
terms, a single photon in the resonator can simultaneously excite
two qubits. Finally, we discuss the quantum interference effects
between four transition paths. Compared with the similar dynamics
studied in Refs.~\cite{Garziano2016,Zhao2017}, the whole
transition process proposed now only involves a \emph{single}
intermediate step, and the process rate can be much faster.
Additionally, our proposal does not require to induce \emph{both}
longitudinal and transverse couplings~\cite{Garziano2016}, so the
superconducting qubit can work at the optimal point and, thus, the
pure-dephasing rate of the qubits can be effectively
suppressed~\cite{Fedorov10,Stern14,Garziano15,Wang16}. Moreover,
we consider \emph{qubits instead of the qutrits} studied in
Ref.~\cite{Zhao2017}. By discussing the parameters in our system,
we find that the coherence rate can easily exceed the decoherence
rate, and it is possible to observe these quantum effects with
current experimental setups.

Superconducting circuits with Josephson qubits are a suitable
platform to explore our proposal as will be discussed in detail in
Sec.~II. We note that the past few years have witnessed the rapid
development in quantum control and quantum engineering based on
superconducting quantum circuits~\cite{Gu2017, Makhlin01, You05,
Liu05sr, rClarke08, DiCarlo10,You2011,rBuluta11,
Xiang13,rGeorgescu14}. The current manufacturing, control, and
detection technologies for the superconducting devices are
mature~\cite{vanderWal2000,Liu05,Neeley2008,Chen11,Inomata16}.
Many quantum phenomena in atomic physics and quantum optics, such
as vacuum Rabi oscillations~\cite{Gu2017}, Autler-Townes
splitting~\cite{Mika09,Wang2016oc,Gu16}, and Fock states
generation~\cite{Hofheinz2008,Hofheinz2009,Premaratne2017}, have
been demonstrated based on superconducting quantum
circuits~\cite{Gu2017}. Moreover, since the dipole moments of a
superconducting qubit are extremely large compared with the ones
in natural atoms, the coupling strength in a circuit QED
system~\cite{rBlais04,Xiang13}, can enter into the strong,
ultrastrong~\cite{Niemczyk2010,FornDiaz2010,Baust2016,FornDiaz2017,Chen17},
and even deep-strong~\cite{Yoshihara2017} regimes. All these
advantages make superconducting quantum circuits an ideal platform
for exploring various quantum effects beyond the RWA.

\section{Model}

\begin{figure}[tb]
\centering \includegraphics[width=8.2cm]{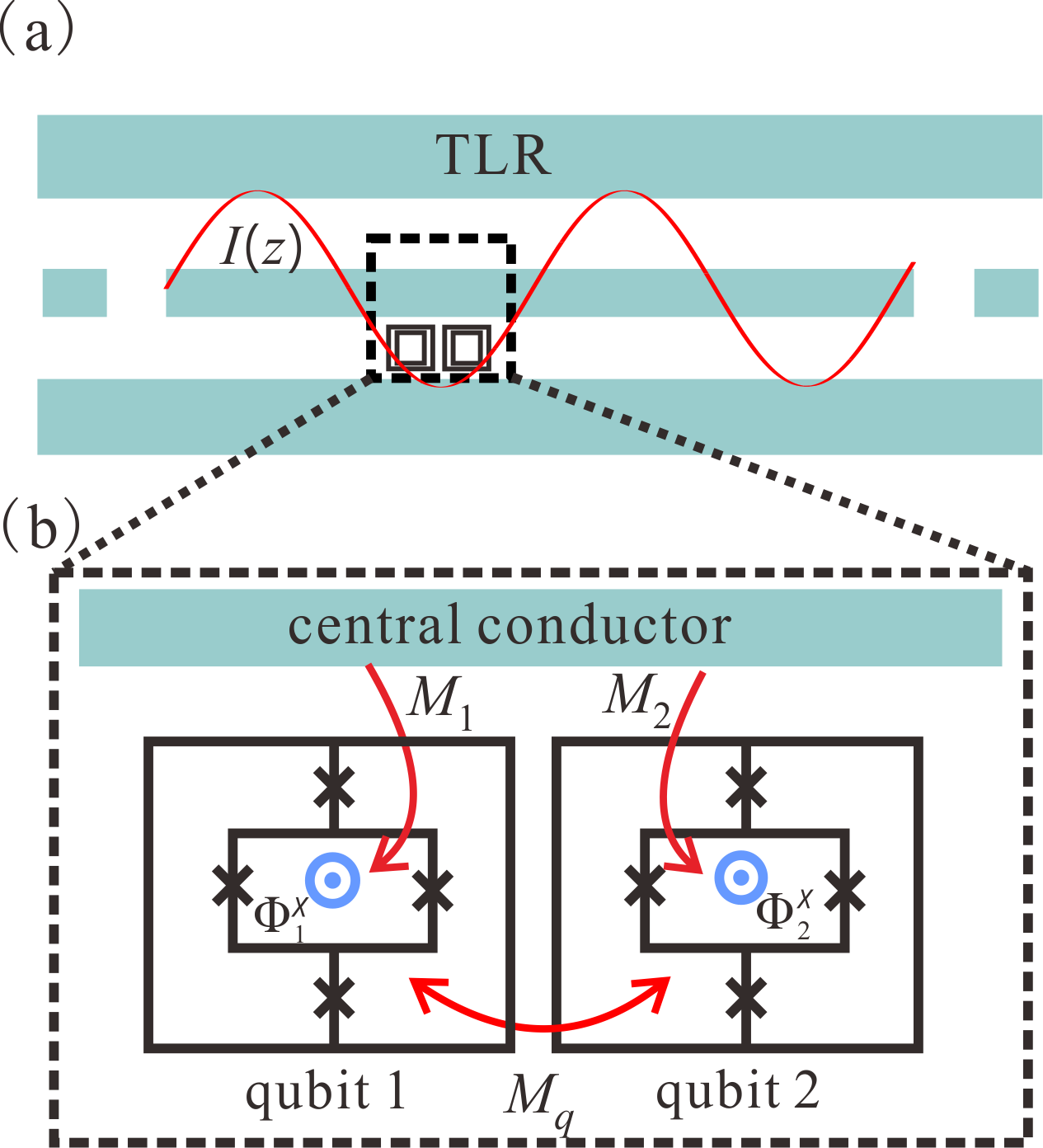} \caption{(Color
online) (a) Schematic circuit layout of our proposal. The central
conductor of the transmission-line resonator (TLR) is stretched in
the $z$ direction, and the two flux qubits are placed at the
antinode positions of the TLR standing-wave current $I(z)$. Here
we assume that each qubit is composed of two symmetric
gradiometric loops and a SQUID loop. (b) The quantized current
$I(z)$ through the TLR central conductor can interact with the
first and second qubit SQUID loops via mutual inductances $M_{1}$
and $M_{2}$, respectively. Each qubit is composed of four
Josephson junctions (modeled by black crosses). To bias the energy
gap of each qubit, a static flux $\Phi_{j}^{x}$ must be through
the $j$th SQUID loop. The two qubits couple together with their
mutual inductance $M_{q}$.} \label{fig1}
\end{figure}
Our model can be implemented in a superconducting quantum circuit
layout with Josephson junctions. As shown in Fig.~\ref{fig1}, we
consider two gap-tunable flux
qubits~\cite{Mooij99,Orlando99,You07,Paauw09,Fedorov10,Stern14}
placed in a transmission-line resonator (TLR), and coupled
together with an antiferromagnetic
interaction~\cite{Niskanen06,Grajcar06,Harris09}. The Hamiltonian
for the two qubits is expressed as (setting $\hbar =1$):
\begin{equation}
\bar{H}_{q}=\frac{1}{2}\sum_{j=1}^{2}(\Delta _{j}\bar{\sigma}%
_{j}^{x}+\epsilon _{j}\bar{\sigma}_{j}^{z})+J\bar{\sigma}_{1}^{z}\bar{%
\sigma}_{2}^{z},  \label{eq1}
\end{equation}%
where the energy basis $\epsilon _{j}=2I_{p,j}(\Phi _{j}^{z}-\Phi
_{0}/2)$ can be controlled via the flux $\Phi _{j}^{z}$ through
the two symmetric gradiometric loops~\cite{Paauw09,Stern14},
$\Delta _{j}$ is the energy gap, $\Phi _{0}$ is the flux quantum,
and $\bar{\sigma}_{z}^{j}$ and $\bar{\sigma}_{x}^{j}$ are the
Pauli operators for the $j$th qubit in the basis of persistent
current states: $|\uparrow_{j}\rangle $ (counterclockwise) and
$|\downarrow_{j}\rangle $ (clockwise) with amplitude
$I_{p,j}$~\cite{Mooij99,Orlando99}. The dipole-dipole interaction
strength $J$ is given by $J=M_{q}I_{p,1}I_{p,2}$, where $M_{q}$ is
the mutual inductance between two
qubits~\cite{Niskanen06,Grajcar06,Harris09}. For such a layout
arrangement, the mutual inductance $M_{q}$ and coupling strength
$J$ are determined by the geometry and spatial relation, when the
qubits are placed next to each other. Alternatively, as discussed
in Refs.~\cite{Averin2003,Wallquist2005,Harris07,Tsomokos10}, one
can achieve a tunable indirect coupling by employing a coupler,
which allows a flexible coupling between two distant qubits. Here
we just assume the two flux qubits as an example, and they can be
replaced by some other type of superconducting artificial qubits
(see, e.g., Figs.~1 and~2 in Ref.~\cite{Gu2017}).

The energy gap $\Delta _{j}$ can be controlled conveniently by
adjusting the static flux $\Phi _{j}^{x}$ through the SQUID loop.
Since the sizes of two qubits ($\sim10~\mu$m) are negligible
compared with the TLR wave-length ($\sim$~cm), we assume that the
resonator current is independent of the resonator position, and
the dipole approximation is valid here. The qubits are placed at
the antinode position $z_{0}$ of the TLR
current~\cite{rBlais04,Xiang13,Gu2017}. Since each qubit has two
symmetric gradiometric loops, the flux contribution from the
current $I(z)$ in the central conductor of the TLR vanishes to the
first order of the energy bias
$\epsilon_{j}$~\cite{Paauw09,Fedorov10,Stern14}. However, the current
$I(z)$ in the central conductor of the TLR can produce flux
perturbations to the SQUID loop of the $j$th qubit via the mutual
inductance $M_{j}$~\cite{Paauw09,Fedorov10}. Therefore, $\Delta
_{j}$ can be expressed as
\begin{equation}
\Delta _{j}=\Delta _{j}(\Phi _{j0}^{x})+R_{j}M_{j}I(z_{0}),
\label{eq2}
\end{equation}%
where $R_{j}=\partial \Delta _{j}(\Phi _{j}^{x})/\partial \Phi
_{j}^{x}$ is the sensitivity of the energy gap $\Delta _{j}$ on
the static-flux frustration at the position $\Phi
_{j0}^{x}$~\cite{Paauw09,Fedorov10}. The quantized current of the
TLR can be directly obtained from the quantization of the voltage
and expressed as~\cite{Yurke84,rBlais04,Gu2017,Clerk10}:
\begin{equation}
I(z_{0})=\sqrt{\frac{\omega }{2L_{0}L}}(a+a^{\dag }),  \label{eq3}
\end{equation}%
where $L_{0}$ is the inductance per unit length of the TLR, $a$
($a^{\dag}$) denotes the annihilation (creation) operator of a
microwave photon in the TLR, $\omega$ is the resonant mode
frequency considered here, and $L$ is the total length of the
resonator~\cite{Yurke84,rBlais04}. The coupling strength between
the $j$th qubit and a single microwave photon in the resonator has
the form
\begin{equation}
g_{j}=R_{j}M_{j}\sqrt{\frac{\omega }{2L_{0}L}},  \label{eq4}
\end{equation}%
and the total Hamiltonian for the whole system can be written
as
\begin{eqnarray}
\bar{H}_{T} &=&\omega a^{\dag}a+\frac{1}{2}\sum_{j=1}^{2}(\Delta _{j}\bar{\sigma%
}_{j}^{x}+\epsilon _{j}\bar{\sigma}_{j}^{z})  \notag \\
&&+\sum_{j=1}^{2}g_{j}\bar{\sigma}_{j}^{x}(a+a^{\dag })+J\bar{\sigma}_{1}^{z}%
\bar{\sigma}_{2}^{z}.   \label{eq5}
\end{eqnarray}
In an experiment, if we apply static fluxes $\Phi _{1}^{x}$ and
$\Phi_{2}^{x}$ through the SQUID loop of two qubits with the
opposite (same) direction, the flux sensitivities of the energy
gaps $R_{1}$ and $R_{2}$ are of the opposite (same) sign.
Moreover, by setting the $j$th qubit working at different
energy-gap points $\Delta_{j}$, the amplitude of $R_{1}$ and
$R_{2}$ can be easily modified~\cite{Paauw09,Fedorov10}. It can be
found that $R_{1}$ and $R_{2}$ can directly determine the
strengths and relative sign between $g_{1}$ and $g_{2}$. The
coupling strengths between the qubits and resonator can be
conveniently adjusted in this circuit QED system according to
Eq.~(\ref{eq4}). In Sec.~IV, we demonstrate how to obtain
different interference effects by modifying $R_{1}$ and $R_{2}$.

To minimize the pure dephasing effect of two qubits induced by the
flux noise, we often operate the qubits at their optimal points
with $\epsilon _{j}=2I_{p,j}(\Phi _{j}^{z}-\Phi _{0}/2)=0$, by
applying a static flux $\Phi _{j}^{z}=\Phi
_{0}/2$~\cite{Paauw09,Stern14} through two gradiometric loops. In
the new basis of the eigenstates $|e_{j}\rangle =(|\uparrow
_{j}\rangle +|\downarrow _{j}\rangle )\sqrt{2}$ and $|g_{j}\rangle
=(|\uparrow _{j}\rangle
-|\downarrow _{j}\rangle )\sqrt{2}$, we can rewrite the Hamiltonian in Eq.~(%
\ref{eq5}) as
\begin{equation}
H_{T}=\omega a^{\dag }a+\frac{1}{2}\sum_{j=1}^{2}\Delta _{j}\sigma
_{j}^{z}+\sum_{j=1}^{2}g_{j}\sigma _{j}^{z}(a+a^{\dag})+J\sigma
_{1}^{x}\sigma _{2}^{x},  \label{eq6}
\end{equation}%
where $\sigma _{j}^{z}=|e_{j}\rangle \langle e_{j}|-|g_{j}\rangle
\langle g_{j}|$ and $\sigma _{j}^{x}=\sigma _{j}^{+}+\sigma
_{j}^{-}=|e_{j}\rangle \langle g_{j}|+|g_{j}\rangle \langle
e_{j}|$. It can be found that the qubit-resonator coupling is
of a longitudinal form, rather than that in the Rabi model
for standard QED systems.

In this paper, we assume that two qubits are nearly resonant,
i.e., $\Delta _{1}\backsimeq \Delta _{2}$, but all our discussions
here can be applied to the case when the two qubits are far off
resonance. The last term in Eq.~(\ref{eq6}) describes the
dipole-dipole interaction between two artificial atoms, which can
be separated into two parts, i.e., the excitation-number
conserving terms
\begin{equation}
  H_{\text{R}}=J(\sigma _{1}^{-}\sigma _{2}^{+}+\text{H.c.}),
 \label{HR}
\end{equation}
and the counter-rotating terms
\begin{equation}
  H_{\text{CR}}=J(\sigma _{1}^{+}\sigma _{2}^{+}+\text{H.c.}).
 \label{HCR}
\end{equation}
It is known that $H_{\text{CR}}$ describes an excitation-number
nonconserving process that two excitations are created
(annihilated) at the same time. This virtual process happens with
an extremely low probability at a rapid oscillating
rate~\cite{Scully1997}. In a conventional analysis of such
dipole-dipole coupling dynamics, the evolution of the two resonant
qubits is dominated by the excitation-number conserving term
$H_{\text{R}}$ before the coupling strength $J$ enters into the
ultrastrong-coupling regime. The counter-rotating term
$H_{\text{CR}}$ is only significant when the coupling reaches the
ultrastrong or deep-strong coupling regimes. However, in this
work, we find the interesting phenomenon that $H_{\text{CR}}$,
rather than $H_{\text{R}}$, dominates the evolution process even
\emph{without} considering the ultrastrong-coupling regime, i.e.,
$\max{\left\{J,g_{i}\right\}}<0.1\times\min{\left\{\omega,\Delta_{j}\right\}}$.

\section{How to observe pure effects of counter-rotating terms}
\begin{figure}[tb]
\centering \includegraphics[width=8.2cm]{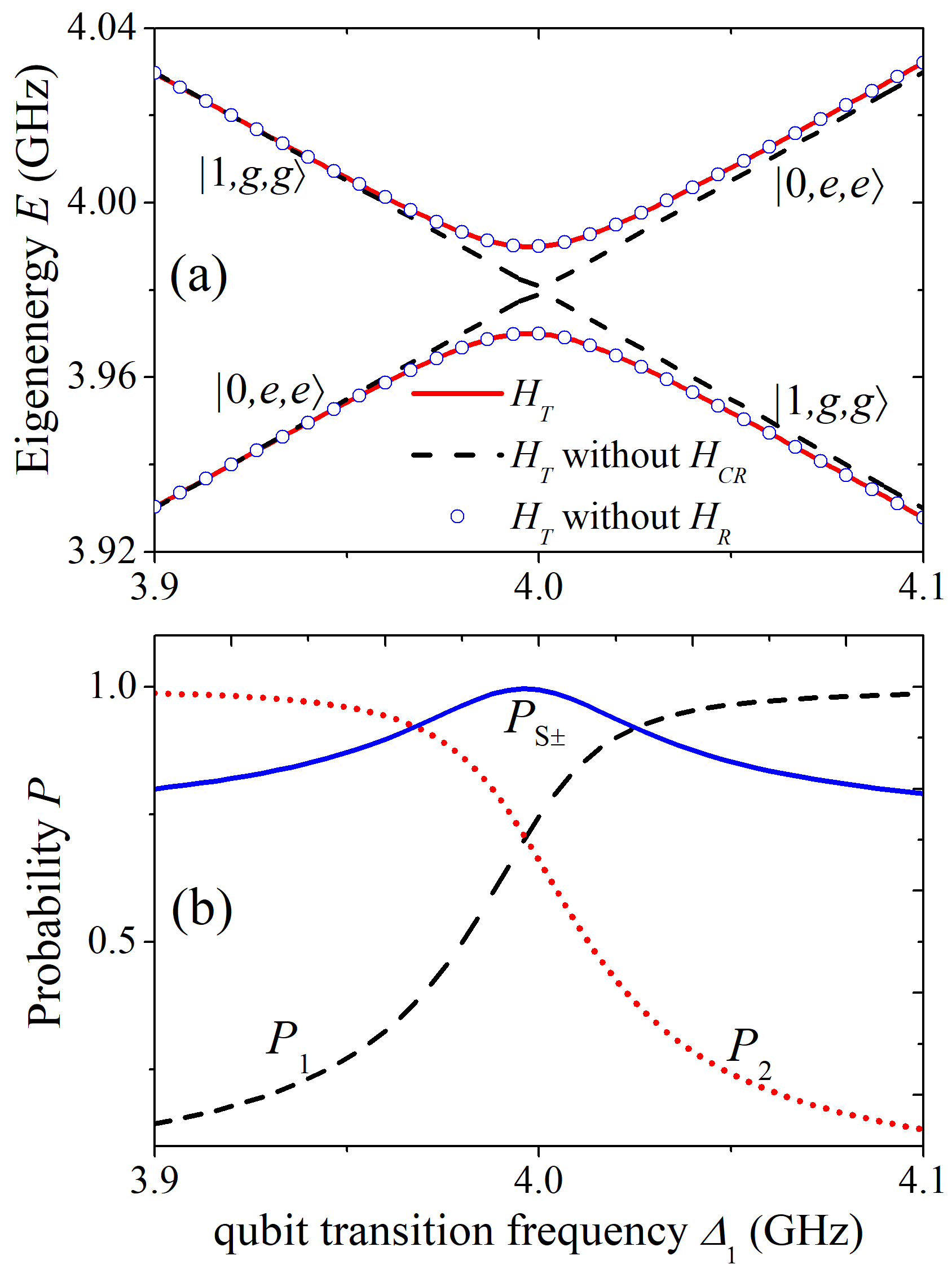} \caption{(Color
online) (a) Eigenenergies $E_{3}$ and $E_{4}$ as functions of the
first qubit transition frequency $\Delta_{1}$. Numerical results
are calculated with the original Hamiltonian $H_{T}$ in
Eq.~(\ref{eq6}) (red solid curve), $H_{T}$ without the
counter-rotating terms $H_{\text{CR}}$ (black dashed curve), and
$H_{T}$ without the resonant terms $H_{\text{R}}$ (blue dots). It
can be clearly seen that the energy spectrum for $E_{3}$ and
$E_{4}$ exhibits an anticrossing point around
$\Delta_{1}=4~\text{GHz}$. The results without $H_{\text{R}}$,
rather than without the counter-rotating terms $H_{\text{CR}}$,
match well with those of the original Hamiltonian $H_{T}$.
(b) The probabilities $P_{1}$ (black dashed curve), $P_{2}$ (red dot
curve) and $P_{s\pm}$ (blue solid curve) which are defined in Sec. III A, as functions of the first
qubit transition frequency $\Delta_{1}$. The parameters used here
are: $\omega=2\Delta_{2}=8~\text{GHz}$,
$g_{1}=g_{2}=0.2~\text{GHz}$, and $J=0.1~\text{GHz}$.}
\label{fig2}
\end{figure}
We are interested in the regime when $\omega\approx\Delta
_{1}+\Delta _{2}$, and assume that the resonator and second-qubit
frequencies are $\omega=2\Delta_{2}=8~\text{GHz}$. Under current
experimental conditions, the coupling strength between a TLR and a
qubit can easily reach the strong-coupling regime (see,
Ref.~\cite{Gu2017} for a recent review), and we assume that
$g_{1}=g_{2}=0.2~\text{GHz}$. According to Ref.~\cite{Majer05}, a
direct inductive coupling strength between the two flux qubits can
be several hundred MHz. In the following discussions, we set
$J=0.1~\text{GHz}$ .

\subsection{Anticrossing point in energy spectra}
In Fig.~\ref{fig2}(a), by changing the first atomic-transition
frequency $\Delta_{1}$, we plot the energy spectrum of the third
and fourth eigenenergies by numerically solving the eigenproblem
$H_{T}|\psi_{n}\rangle=E_{n}|\psi_{n}\rangle$, with $n=3,4$. It
can be seen that the two energy levels exhibit anticrossing with a
splitting around $\Delta _{1}=4~\text{GHz}$ (red solid curves),
which indicates that there might be two states coupled resonantly.
Specifically, if the counter-rotating terms $H_{\text{CR}}$ in
Eq.~(\ref{eq6}) are neglected, the anticrossing point disappears
(see the dashed black curves). However, without the two-qubit
resonant coupling terms $H_{\text{R}}$, the energy spectrum (blue
dot curves) for the third and fourth eigenstates coincides with
the full Hamiltonian case, which indicates that the resonant
coupling is due to the counter-rotating terms $H_{\text{CR}}$, and
has no relation with $H_{\text{R}}$. We note that the predicted
level anti-crossing is analogous to that observed in the experiment
of Niemczyk et al.~\cite{Niemczyk2010} and other experiments
in the USC regime using superconducting quantum circuits
(for a very recent review see~\cite{Gu2017} and references therein).
Analogous to our model, the emergence of this level anticrossing needs
qubit-oscillator longitudinal couplings. However, as discussed in
Refs.~\cite{Niemczyk2010,Garziano15,Wang16}, the origin of this phenomenon
is due to multi-excitation processes and has a close relation to
both the counter-rotating terms and JC terms in the transverse coupling.
In our proposal, only the counter-rotating
terms contribute to the energy level anticrossing, even far below the USC regime.

In Fig.~\ref{fig2}(b), we plot the probabilities
$P_{1}=|\langle0,e,e|\psi_{4}\rangle|^{2}$,
$P_{2}=|\langle1,g,g|\psi_{4}\rangle|^{2}$, $P_{s+}=|\langle
S_{+}|\psi_{3}\rangle|^{2}$ and $P_{s-}=|\langle
S_{-}|\psi_{4}\rangle|^{2}$(where
$|S_{\pm}\rangle=(|0,e,e\rangle\pm |1,g,g\rangle)/\sqrt{2}$),
changing with $\Delta_{1}$. It can be seen that
$|\psi_{3}\rangle\simeq|1,g,g\rangle$
($|\psi_{4}\rangle\simeq|0,e,e\rangle$) when
$\Delta_{1}\sim 3.9~\text{GHz}$ ($\Delta_{1}\sim 4.1~\text{GHz}$).
Around the anticrossing point, $|S_{+}\rangle\simeq|\psi_{3}\rangle$ and
$|S_{-}\rangle\simeq|\psi_{4}\rangle$.
One may wonder why we are not showing in Fig.~\ref{fig2} the corresponding
plots for the probabilities
$P'_{1}=|\langle1,g,g|\psi_{3}\rangle|^{2}$,
$P'_{2}=|\langle0,e,e|\psi_{3}\rangle|^{2}$, $P'_{1}\approx P_{1}$, and $P'_{2}\approx P_{2}$,
such that we would not see any
differences between the corresponding curves on the scale of Fig.~\ref{fig2}.
Therefore, we can conclude
that the anticrossing point is due to the resonant coupling
between the states $|0,e,e\rangle$ and $|1,g,g\rangle$. The
coherent transfer between these two states corresponds to the same
interesting process discussed in Ref.~\cite{Garziano2016}: that a
single photon in a cavity can excite two atoms simultaneously. One
may wonder how this process can happen in our system with only
longitudinal coupling. To show this, hereafter, we analytically
derive its effective Hamiltonian.

\begin{figure}[tb]
\centering \includegraphics[width=8.4cm]{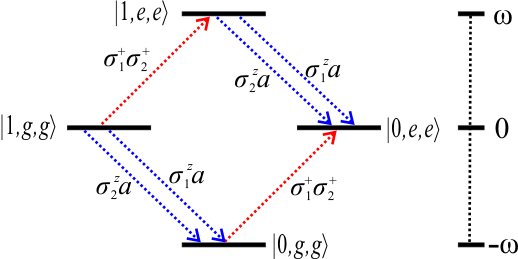} \caption{(Color
online) Sketch of the high-order transitions between the states
$|1,g,g\rangle$ and $|0,e,e\rangle$. The four transition paths,
allowed by the longitudinal couplings (blue arrows) $\sigma
_{j}^{z}a$ and the counter-rotating terms $\sigma _{1}^{+}\sigma
_{2}^{+}$ (red arrows), are mediated by the two states
$|0,g,g\rangle$ and $|1,e,e\rangle$. It is seen that the only
difference between the two paths (with the same intermediate
state) corresponds to different longitudinal couplings, which
results in annihilating a virtual photon via the first (at the
rate $\beta_{1}J$) or second (at the rate $\beta_{2}J$) qubits.}
\label{fig3}
\end{figure}

\subsection{Effective Hamiltonian for the tripartite interaction}
We first perform the polariton transformation of the Hamiltonian
$H_{\text{tot}}$ in Eq.~(\ref{eq6}) given by
\begin{equation}
H_{s1}=e^{S}H_{T}e^{-S},
\end{equation}%
with $S=\sum_{j=1}^{2}\beta _{j}\sigma _{j}^{z}(a^{\dag }-a)$,
where $\beta _{j}=g_{j}/\omega$ is the Lamb-Dicke
parameter for the $j$th qubit-resonator longitudinal coupling.
Thus, we obtain
\begin{eqnarray}
H_{s1}&=&\omega a^{\dag }a+\frac{1}{2}\sum_{j=1}^{2}\Delta_{j}\sigma
_{j}^{z}-\chi\sigma_{1}^{z}\sigma_{2}^{z}   \notag  \\
&&+J\prod_{j=1}^{2}[\sigma _{j}^{+}e^{2\beta _{j}(a^{\dag }-a)}+\text{%
H.c.}], \label{eq7}
\end{eqnarray}%
where $\chi=4g_{1}g_{2}/\omega$ is the $\sigma_{1}^{z}\sigma_{2}^{z}$ coupling strength
between two qubits. Given that $\beta _{j}\ll 1$ ($\beta _{1}=\beta _{2}=0.025$), the
last term in Eq.~(\ref{eq7}) can be expanded to first order in
$\beta_{j}$. Therefore, $H_{s1}$ can be approximately written as
\begin{eqnarray}
H_{s2}&\simeq&\omega a^{\dag }a+\frac{1}{2}\sum_{j=1}^{2}\Delta_{j}\sigma
_{j}^{z}-\chi\sigma_{1}^{z}\sigma_{2}^{z} \notag \\
&+&J\big[(\sigma _{1}^{+}+\sigma _{1}^{-})+2\beta _{1}(\sigma _{1}^{+}-\sigma _{1}^{-})(a^{\dag}-a)\big] \notag \\
&\times&\big[(\sigma _{2}^{+}+\sigma _{2}^{-})+2\beta _{2}(\sigma _{2}^{+}-\sigma _{2}^{-})(a^{\dag}-a)\big]. \label{eq8}
\end{eqnarray}
The last term describes various types of multi-excitation
interactions among the qubits and the field, such as
$\sigma _{1}^{+}\sigma _{2}^{+}a$ and $\sigma _{1}^{+}\sigma _{2}^{+}a^{2}$.
To observe the effects of the counter-rotating terms in the
dipole-dipole coupling, here we assume that the dipole-dipole
coupling $J\ll\rm{min}\{{\Delta _{j}, \Delta _{j}\pm\omega}\}$
($j=1,2$) and $\omega=\Delta _{1}+\Delta _{2}$. Employing the commutation relations
$[\sigma_{1}^{z}\sigma_{2}^{z},\sigma_{1}^{\pm}\sigma_{2}^{\pm}]=0$, and applying the
unitary transformation
\begin{equation}
U=\exp\Big[-i\Big(\omega a^{\dag }a+\frac{1}{2}\sum_{j=1}^{2}\Delta _{j}\sigma
_{j}^{z}-\chi\sigma_{1}^{z}\sigma_{2}^{z}\Big)t\Big]
 \label{U}
\end{equation}
to the Hamiltonian in Eq.~(\ref{eq8}) for the time $t$,
 we obtain the resonant
Hamiltonian by neglecting the rapidly oscillating terms
\begin{equation}
H_{\rm{eff}}=G_{s}(a\sigma _{1}^{+}\sigma _{2}^{+}+a^{\dag}\sigma
_{1}^{-}\sigma _{2}^{-}), \label{eq9}
\end{equation}
with the effective coupling strength
\begin{equation}
  G_{s}=2J(\beta _{1}+\beta
_{2})=\frac{2J(g_{1}+g_{2})}{\omega}.
 \label{Gs}
\end{equation}
We can clearly find that Eq.~(\ref{eq9}) describes the energy of a
photon in a resonator splitting into two parts and simultaneously
exciting two qubits. In the original Hamiltonian in
Eq.~(\ref{eq6}), the longitudinal coupling between the $j$th qubit
and the resonator, i.e., $\sigma _{j}^{z}a^{\dag}$ ($\sigma
_{j}^{z}a$) corresponds to the creation (annihilation) of a
virtual photon in the resonator at a rapid rate $\omega$. The
counter-rotating term in the qubit coupling, i.e., $\sigma
_{1}^{+}\sigma _{2}^{+}$ ($\sigma _{1}^{-}\sigma _{2}^{-}$),
describes the process of simultaneously exciting
(de-exciting) two qubits. This term does not conserve the
excitation number, and is also a virtual process oscillating at a
high frequency $(\Delta _{1}+\Delta _{2})$. However, as shown in
Fig.~\ref{fig3}, these excitation-number-nonconserving processes
can be combined together to form four resonant transition
processes. The coherent-transfer rate between the states
$|n+1,g,g\rangle$ and $|n,e,e\rangle$ is $\sqrt{n+1}G_{s}$, with
$|n,g,g\rangle$ and $|n+1,e,e\rangle$ being two intermediate
states, respectively. In contrast to conventional QED problems,
where we neglect the counter-rotating terms, here $H_{\text{CR}}$
plays a key role in exciting the two qubits simultaneously, while
the resonant terms $H_{\text{R}}$ have no effect. Therefore, the
RWA is not valid here, even if the couplings are not in the
ultrastrong-coupling regime.

By assuming the same parameters as in Fig.~\ref{fig2} and
$\Delta=4~\rm{GHz}$, the effective coupling strength can be as
strong as $G_{s}=10~\text{MHz}$. Compared with the results in
Refs.~\cite{Garziano2016,Zhao2017}, there is only a \emph{single}
(rather than \emph{two}) intermediate virtual state
$|0,g,g\rangle$ during the process where a single photon excites
two atoms. Consequently, the corresponding coupling rates are
faster by about one order of magnitude.
\subsection{Adiabatic Landau-Zener transition}
In the vicinity of the anticrossing point, we first examine the
adiabatic Landau-Zener transition
effect~\cite{Zener696,Rubbmark81,Shevchenko2010} without
considering the dissipative channels. Assuming that the atomic
transition frequency $\Delta_{1}$ is linearly dependent in time,
i.e.,
\begin{equation}
\Delta_{1}(t)=\Delta_{1}(0)+vt, \label{eq10}
\end{equation}
where $\Delta_{1}(t)$ sweeps through the anticrossing point at a
velocity $v$. In an experiment, it is convenient to tune
$\Delta_{1}(t)$ linearly by changing the flux $\Phi_{x,1}$ through
the SQUID loop. We assume that the system is initially in its
fourth eigenstate $|\psi_{4}\rangle\simeq|1,g,g\rangle$. When
changing $\Delta_{1}(t)$ linearly, the system might jump to the
lower eigenstate $|\psi_{3}\rangle$ due to the \emph{diabatic}
transition. In other words, this means that the system evolves far
away from a quasi-steady state and transitions between different
eigenstates can occur. Thus, the final transition probability to
the state $|\psi_{3}\rangle\simeq|1,g,g\rangle$
($\Delta>4~\rm{GHz}$) can be approximately expressed by the
Landau-Zener formula~\cite{Zener696,Rubbmark81,Ma15}, i.e.,
\begin{equation}
P_{\psi_{3}}=\exp\left[{-2\pi\frac{G_{s}^2}{dE_{\Delta}/dt}}\right],
\label{eq11}
\end{equation}
where $E_{\Delta}=E_{4}(t)-E_{3}(t)$ is the eigenenergy difference
between the fourth and third eigenstates, and $dE_{\Delta}/dt$ is
the sweeping rate. Here we simply have $dE_{\Delta}/dt\simeq v$.
If the energy-sweeping speed $v$ is extremely slow and satisfies
the relation $2\pi G_{s}^2\gg v$, the anticrossing point traverses
\emph{adiabatically}~\cite{Rubbmark81,Ma15}. In this case, the
system approximately evolves along the fourth-energy curve, and
the system rarely jumps to the third eigenstate after the
sweeping, i.e., $P_{\psi_{3}}\ll1$.
\begin{figure}[tb]
\centering \includegraphics[width=7.8cm]{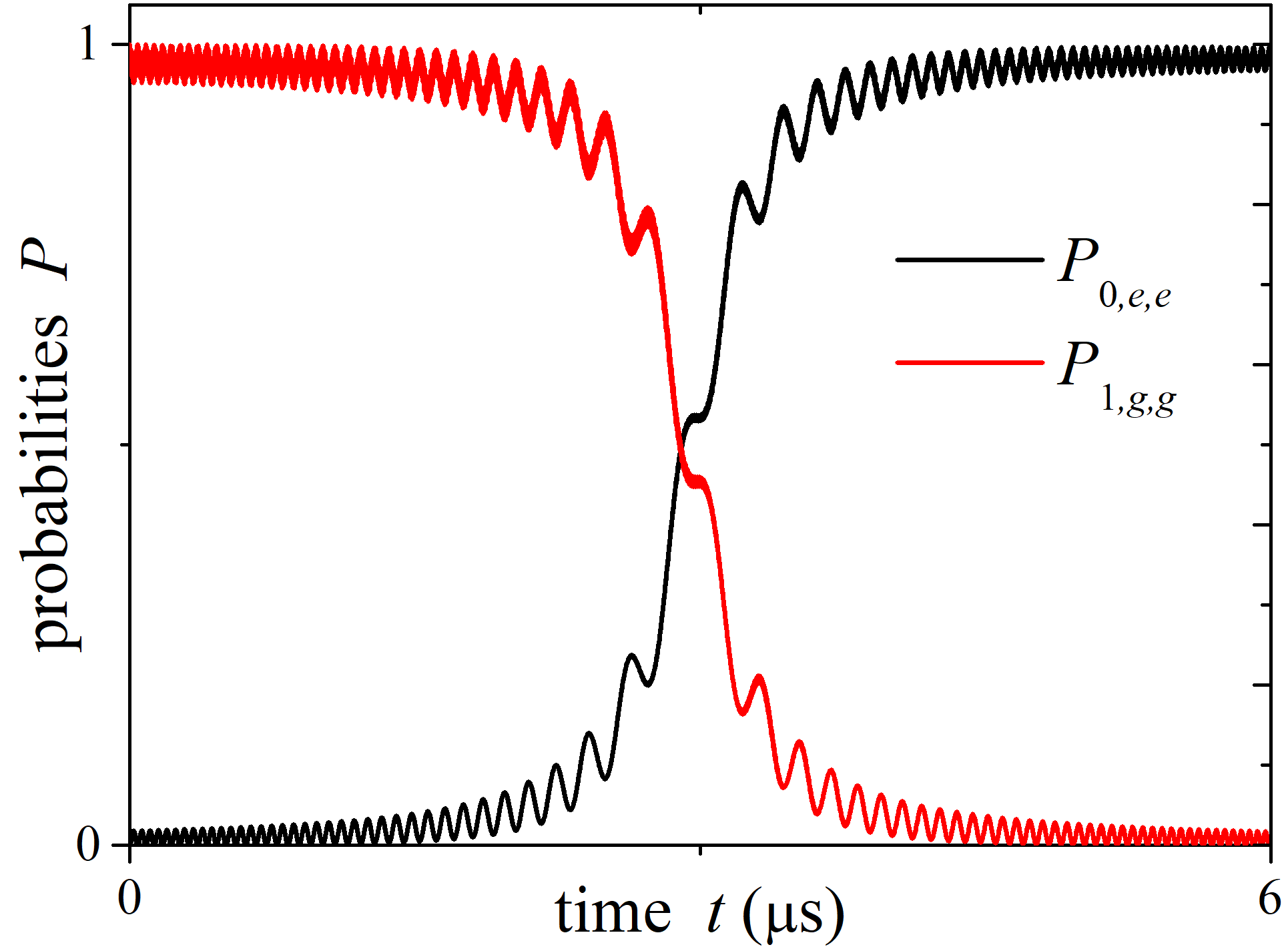} \caption{(Color
online) The time-dependent Landau-Zener transition process is
achieved by slowly changing the first-qubit frequency $\Delta_{1}$
with a sweeping rate $v=6\times10^{-5}~(\text{GHz})^{2}$. The
initial state is $|1,g,g\rangle$ (red curve). Due to the resonant
coupling effects, the probability of the initial state is
gradually decreased, and the system is adiabatically transferred
into the state $|0,e,e\rangle$ (black curve) with the final
probability $P_{1,e,e}\simeq0.99$. The other parameters used here are
the same as those in Fig.~\ref{fig2}(a).} \label{fig4}
\end{figure}

In Fig.~\ref{fig4}, by setting $\Delta_{0}(t)=3.84~\text{GHz}$ and
$v=6\times10^{-5}~(\text{GHz})^{2}$, we numerically simulate the
evolution dominated by the Schr\"{o}dinger equation, and plot the
probabilities of the states $|1,g,g\rangle$ and $|0,e,e\rangle$
changing with time, respectively. It can be clearly seen that the
probability $P_{0,e,e}$ gradually increases from 0 to $\sim0.99$.
The transition time is of the order of several microseconds. For
the final states, there is still a low probability $P_{1,g,g}$
because of the extremely-weak diabatic-transition
effect~\cite{Rubbmark81}. During this process, the excitation
energy of a single photon is split into two parts, to effectively
excite the two flux qubits. The dynamics of this Landau-Zener
transition provides a strong evidence of the resonant coupling
between the states $|1,g,g\rangle$ and $|0,e,e\rangle$.
\section{Quantum Rabi oscillations and interference effects between four transition paths}
To examine the deterministic transition between the states
$|1,g,g\rangle$ and $|0,e,e\rangle$, the rate of the adiabatic
Landau-Zener transition process is extremely slow. Therefore, we
can simply observe the Rabi oscillation between these two states.

We assume that the resonator and two qubits are coupled to
the vacuum environment and the initial states of the system are
their ground states $|0,g,g\rangle$. The coherent electromagnetic
field is applied via a 1D transmission line, which couples to one
side of the resonator via a capacitance~\cite{Underwood12}.

We can inject a single photon into the resonator by applying a
Gaussian pulse, i.e., to prepare the initial state as
$|1,g,g\rangle$, and the corresponding drive has the form
\begin{equation}
H_{\rm drv}(t)=
A\frac{\exp{\left[-(t-t_{0})^{2}/(2\tau^{2})\right]}}{\sqrt{2\pi}\tau}\left(ae^{i\omega
t}+a^{\dag}e^{-i\omega t}\right). \label{eq12}
\end{equation}
where $A$, $t_{0}$, and $\tau$ are the amplitude, central-peak
position, and width of a Gaussian pulse.  However, for a resonator
without nonlinearity, the higher-energy states (for example,
$|2,g,g\rangle$) can also be effectively populated. We can employ
an ancillary superconducting qubit to induce some nonlinearities
of the resonator with a Kerr-type Hamiltonian
$H_{\text{Kerr}}=\chi_{3} a^{\dag2}a^{2}$~\cite{Hoffman11,Garziano2016}. Here, $\chi_{3}$ is the
effective Kerr-interaction strength proportional to third-order
susceptibility.  As a result, the Hamiltonian for the whole system
can be written as
\begin{equation}
H_{t}=H_{T}+H_{\text{Kerr}}+H_{\rm drv}(t). \label{eq12b}
\end{equation}

\subsection{Modified input-output relation}
In standard QED systems, the output and correlation signal are
obtained via photodetection methods. As discussed in
Refs.~\cite{Ridolfo12,Garziano13,Ridolfo13,Garziano15}, when the
coupling is in the strong or ultrastrong coupling regimes, the
eigenstates of the system are the highly-dressed states which are
different from the bare eigenstates of the resonator and qubits,
and the standard input-output relation fails to describe the
output field. For example, the output field photon flux is not
proportional to the conventional first-order correlation functions
of the cavity operators any more~\cite{Ridolfo12}. By contrast to
this, the output field from the cavity is linked to the
electric-field operator $X=a+a^{\dag}$ (rather than the
annihilation operator $a$)~\cite{Ridolfo12,Collett84,Gardiner85}.

To discuss problems more explicitly and consider more general
cases, we employ the modified formula of the input-output relation
and correlation functions in the following discussions. Defining
the positive and negative frequency~\cite{Garziano13} components
of the operator $X$ as
\begin{equation}
X^{+}=\sum_{j,k>j}X_{jk}|\psi_{j}\rangle\langle\psi_{k}|,\quad
X^{-}=(X^{+})^{\dag}, \label{eqoperator1}
\end{equation}
where $X_{jk}=\langle\psi_{j}|(a+a^{\dag})|\psi_{k}\rangle$, the
modified input-output relation under the Markov approximation
can be reexpressed as
\begin{equation}
A_{\text{out}}=A_{\text{in}}-\sqrt{\kappa}X^{+}, \label{eq14}
\end{equation}
where $A_{\text{in}}$ is the input vacuum
noise~\cite{Ridolfo12,Collett84,Gardiner85}, $\kappa$ is the
photon escape rate from the resonator~\cite{Underwood12}, and
$A_{\text{out}}$ is the output field operator of the
form~\cite{Garziano13}:
\begin{equation}
A_{\text{out}}(t)=\frac{1}{2\sqrt{\pi\omega\upsilon}}\int_{0}^{\infty}d\omega^{\prime}a^{\prime}(\omega^{\prime},t_{1})e^{-i\omega^{\prime}(t-t_{1})}+\rm{H.c.},
\label{outfield}
\end{equation}
where $\upsilon$ is the phase velocity of the mode $\omega$,
$a^{\prime}$ is the annihilation operator of the continuous mode
with frequency $\omega^{\prime}$ outside the resonator. The
output photon flux can be expressed as $\Theta=\kappa\langle
X^{-}X^{+}\rangle$.

\subsection{Rabi oscillations based on numerically simulating the master equation}
Under the Born-Markov approximation and assuming that the
resonator and the qubits are coupled to the zero-temperature
vacuum reservoir, the evolution for the system can be described by
the master equation of the Lindblad form~\cite{Ridolfo13,Ma15}
\begin{equation}
\frac{d\rho(t)}{dt}=-i[H_{t},\rho(t)]+\kappa
D[X^{+}]\rho(t)+\sum_{j=1,2}\Gamma_{j}D[C_{j}^{+}]\rho(t),
\label{eq15}
\end{equation}
where
$D[O]\rho(t)=[2O\rho(t)O^{\dag}-O^{\dag}O\rho(t)-\rho(t)O^{\dag}O]/2$
is the Lindblad superoperator, and $\Gamma_{j}$ is the decay rate
of the $j$th qubit. Our proposal requires \emph{only longitudinal}
couplings between the qubits and resonator, rather than
\emph{both} longitudinal and transverse couplings used in
Ref.~\cite{Garziano2016}. Thus, the flux qubits could now work at
their optimal points, and the pure-dephasing rates induced by flux
noise, can be minimized, as discussed in
Refs.~\cite{Paauw09,Fedorov10,Stern14}. The coherence time of a
flux qubit can be of several $\mu\rm{s}$. Here we assume that
$\Gamma_{1}=\Gamma_{2}=0.2~\text{MHz}$. In an experiment, a
superconducting resonator with quality factor over $10^{4}$ can be
easily fabricated~\cite{Megrant12}. We consider the decay rate of
the resonator to be $\kappa=0.4~\text{MHz}$ ($Q=2\times10^{4}$).
Therefore, under current experimental approaches, the
coherent-transition rate $G_{s}$ can easily overwhelm all the
decoherence channels in our proposal.

The emission field for the $j$th qubit is proportional to the
zero-time delay correlation function $\langle
C^{-}C^{+}\rangle$~\cite{Garziano15}, where
\begin{equation}
C_{j}^{+}=\sum_{i,k>i}C_{j,ik}|\psi_{i}\rangle\langle\psi_{k}|,\quad
C_{j}^{-}=(C_{j}^{+})^{\dag}, \label{eqoperator2}
\end{equation}
with the coefficients $C_{j,ik}=\langle\psi_{i}|(\sigma
_{+}^{j}+\sigma _{-}^{j})|\psi_{k}\rangle$. It can be clearly
found that the emission operator is also divided into positive and
negative frequency parts. The zero-delay two-qubit correlation
function $$G_{q}^{(2)}(0)=\langle
C_{1}^{-}C_{2}^{-}C_{2}^{+}C_{1}^{+}\rangle$$ is proportional to
the probability that two qubits are both in their exited
states~\cite{Garziano2016}.

\begin{figure}[tb]
\centering \includegraphics[width=8.2cm]{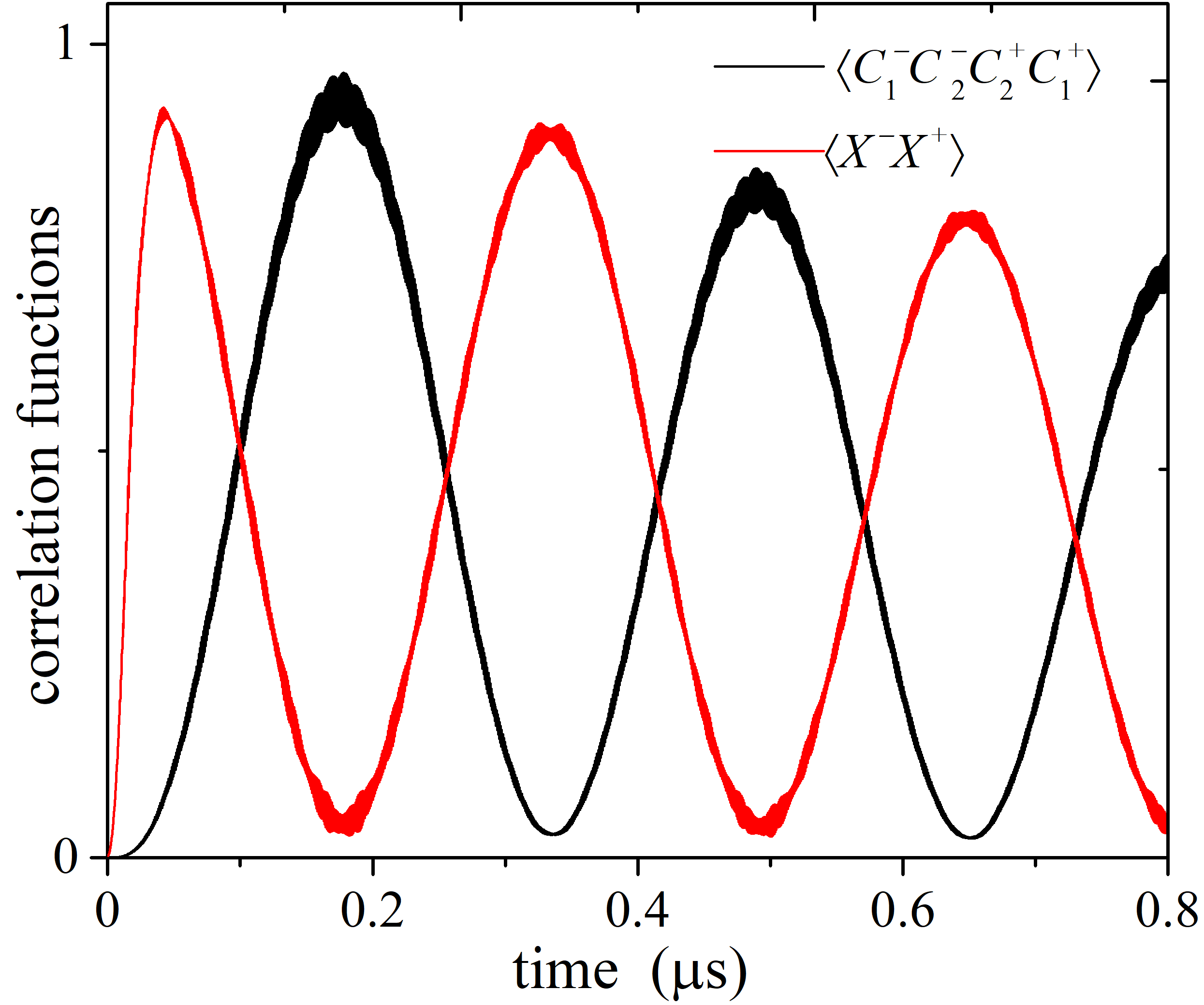} \caption{(Color
online) The intra-resonator photon number $\langle
X^{-}X^{+}\rangle$ (red curve) and zero-delay two-qubit
correlation function $G_{q}^{(2)}(0)$ (black curve) versus time
based on numerically solving the master equation~(\ref{eq15}) with
the decay rates $\Gamma_{1}=\Gamma_{2}=0.2~\rm{MHz}$ and
$\kappa=0.4~\rm{MHz}$. The Kerr nonlinearity is assumed to be
$\chi_{3}=120~\rm{MHz}$. The initial state is $|0,g,g\rangle$. The
Gaussian pulse parameters are: $t_{0}=0$, $\tau=0.02~\mu\rm{s}$,
and $A/(\sqrt{2\pi}\tau)=50~\rm{MHz}$. Other parameters are the
same as those in Fig.~\ref{fig2}(a) with $\Delta_{1}=4~\rm{GHz}$.}
\label{fig5}
\end{figure}

In Fig.~\ref{fig5}, we numerically calculate the photon number
$\langle X^{-}X^{+}\rangle$ inside the resonator and the two-qubit
correlation function $G_{q}^{(2)}(0)$ changing with time. It can
be seen that, due to the Kerr-type nonlinearity, a Gaussian pulse
can create a single photon in the resonator, and the photon flux
can increase rapidly. When $t\gg\tau$, the pump effect of the
Gaussian pulse almost vanishes, and excitation energies can be
coherently transferred between the resonator and two qubits via
the Rabi-oscillation process. Around $t\simeq0.18~\mu\text{s}$,
the two-qubit correlation function $G_{q}^{(2)}(0)$ reaches its
highest value $\sim0.96$, indicating that the two qubits are
strongly correlated and both approximately in their exited states.
Meanwhile the photon number $\langle X^{-}X^{+}\rangle$ reaches it
lowest value, and the injected single photon is effectively
converted into the excitations of two qubits. The reversible
evolution between $G_{q}^{(2)}(0)$ and $\langle X^{-}X^{+}\rangle$
is due to the vacuum Rabi oscillations between the states
$|1,g,g\rangle$ and $|0,e,e\rangle$. Of course, the amplitude of
the oscillations gradually decreases due to the energy-decay
channels.

\subsection{Quantum interference between four transition paths}

\begin{figure}[tb]
\centering \includegraphics[width=8.4cm]{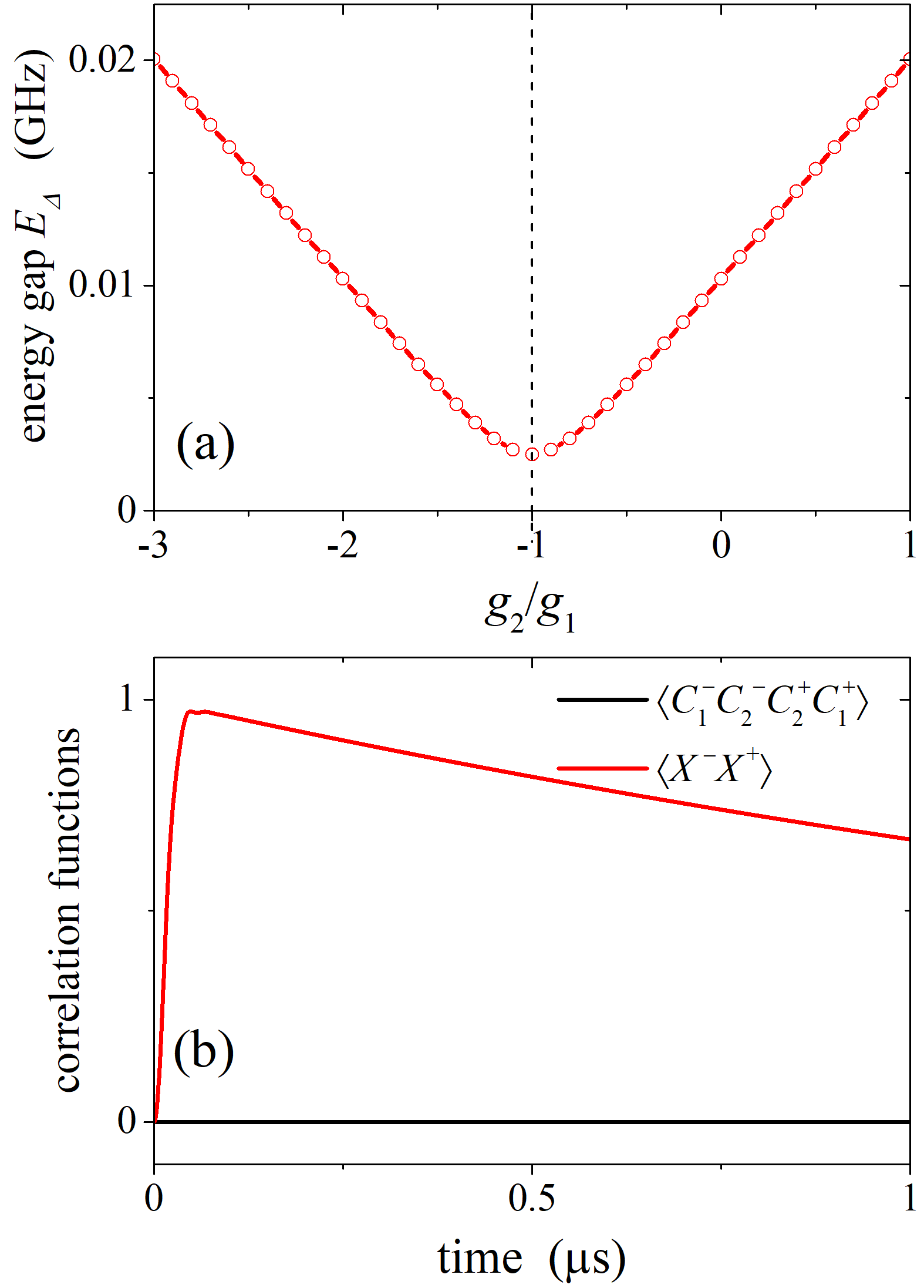} \caption{(Color
online) (a) The energy gap (difference) between the third and
fourth eigenenergies, $E_{\Delta}=E_{4}-E_{3}$, versus the
relative coupling strength $g_{2}/g_{1}$
($g_{1}=0.20~\text{GHz}$). At $g_{2}/g_{1}=-1$ (vertical dashed
line), the gap almost vanishes. (b) Time evolutions of the photon
number $\langle X^{-}X^{+}\rangle$ (red curve) and two-qubit
correlation function $G_{q}^{(2)}(0)$ (black curve), which always
vanishes. Here we set $g_{1}=-g_{2}=0.20~\text{GHz}.$ The Rabi
oscillation disappears here due the destructive interference
effect between the transition paths. Other parameters are the same
as those in Fig.~\ref{fig5}.} \label{fig6}
\end{figure}

Finally, we discuss another interesting phenomenon. As shown in
Fig.~\ref{fig3}, we can find that, for the two paths with the same
intermediate state, the only difference between these paths
corresponds to different longitudinal couplings, which lead to
creating a virtual photon either via the first qubit
($\sigma_{1}^{z}a^{\dag}$) or the second qubit
($\sigma_{2}^{z}a^{\dag}$). The rates of the two paths are
$G_{1}=J\beta_{1}$ and $G_{2}=J\beta_{2}$,
respectively. The coherent transitions between the initial and
final states can be viewed as the interference effect between
these paths, i.e., $G_{s}=2(G_{1}+G_{2})$. As discussed in Sec.
II, the sign and amplitude of $g_{j}$ can be easily tuned by
changing the flux bias direction and the working position of the
energy gap. If $g_{2}$ has opposite sign (i,e., with a $\pi$-phase
difference) but the same amplitude with $g_{1}$, the paths become
destructive, and the coherent transition between the states
$|1,g,g\rangle$ and $|0,e,e\rangle$ vanishes. In
Fig.~\ref{fig6}(a), we plot the anticrossing gap
$E_{\Delta}=E_{4}-E_{3}$ between the third and fourth
eigenenergies changing with the relative strength $g_{2}/g_{1}$.
It can be clearly seen that $E_{\Delta}$ has a dip (almost zero)
at $g_{2}/g_{1}=-1$, indicating that the
anticrossing point almost disappears. Note that $E_{\Delta}$
cannot be exactly equal to zero due to higher-order processes. At
this point, the states $|1,g,g\rangle$ and $|0,e,e\rangle$
decouple from each other. When $g_{2}/g_{1}>0$, the anticrossing
gap $E_{\Delta}$ increases with $g_{2}$ and the transition paths
become constructive.

To observe more clearly the quantum destructive effects between
these paths, we plot the time-dependent evolution of the photon
number $\langle X^{-}X^{+}\rangle$ and the two-qubit correlation
function $G_{q}^{(2)}(0)$. Here we employ the parameters at the
dip in Fig.~\ref{fig6}(a), i.e., $g_{1}=-g_{2}=0.20~\rm{GHz}$. As
shown in Fig.~\ref{fig6}(b), the energy cannot be transferred
between the resonator and two qubits any more, which is different
from the Rabi oscillation in Fig.~\ref{fig5}. Consequently, a
single photon, which is excited by a Gaussian pulse, decays to the
vacuum environment (red curve), and the two-qubit correlation
function $G_{q}^{(2)}(0)$ is always zero (black curves). In such
conditions, the coherent transfer between the states
$|1,g,g\rangle$ and $|0,e,e\rangle$ vanishes due to the
destructive effect, and the counter-rotating-term effect (that a
single photon excites two qubits simultaneously) cannot be
observed. Therefore, in an experiment, we can simply change the
\emph{relative sign and amplitude} of the flux sensitivity $R_{j}$
to observe either destructive or constructive interference effects
caused by the counter-rotating terms.

\section{Discussion and conclusions}

In this paper, we have investigated pure effects of the
counter-rotating terms in the dipole-dipole coupling between two
superconducting qubits. The theoretical analysis shows that when
these two qubits are longitudinally coupled with the same
resonator, the energy of a single photon can effectively excite
two qubits simultaneously. By discussing the anticrossing points
around the resonant regime, we find that this coherent transition
process results from the counter-rotating terms, and has no
relation to the resonant coupling terms between two qubits. In
fact, our results throughout this paper show that when dealing
with a QED system containing longitudinal couplings, we should
examine the energy spectrum of the system carefully before
adopting the standard RWA. The counter-rotating terms might play
an important role in the physical dynamics of the whole system.

Moreover, we have demonstrated the Landau-Zener transition effects
and Rabi oscillations between the states $|1,g,g\rangle$ and
$|0,e,e\rangle$, which are clear signatures of the resonant
coupling between these two states. The energy of a single photon
can be divided to simultaneously excite two qubits via the
\emph{longitudinal} couplings and the counter-rotating terms.
Moreover, this process is combined with four transition paths, and
there can be quantum interference between these paths. We
discussed how to control the system to achieve either destructive
or constructive interference effects. By discussing the
experimentally feasible parameters, we find it is possible to
implement our proposal and observe these quantum effects based on
current state-of-the-art circuit-QED systems.

In fact, if we consider a more general case with
$n\omega=\Delta_{1}+\Delta_{2}$ when deriving the resonant terms
in Eq.~(\ref{eq8}), we can expand this formula to its $n$th order.
In such conditions, a more general resonant Hamiltonian
\begin{equation}
  H^{(n)}_{\rm{eff}}=G_{s}^{(n)}[a^{n}\sigma _{1}^{+}\sigma
_{2}^{+}+(a^{\dag})^n\sigma _{1}^{-}\sigma _{2}^{-}],
 \label{Hn}
\end{equation}
which describes higher-order effects when $n$ photons excite two
qubits simultaneously, might produce observable quantum effects.
However, we should note that the effective rate $G_{s}^{(n)}$
decreases quickly with increasing $n$, which might be
overwhelmed by the nonresonant-oscillating terms and decoherence
processes.

We should emphasize that our proposal here can be a convenient
platform to observe pure quantum effects of the counter-rotating
terms. As we discussed above, these high-order transitions only
contain a single intermediate state, and the rate is much faster
compared with the proposals in Refs.~\cite{Garziano2016,Zhao2017}.
Therefore, the tripartite interaction in Eq.~(\ref{eq9}) provides
a novel way to prepare a type of Greenberger-Horne-Zeilinger
(GHZ) state~\cite{Greenberger90},
$(|1,g,g\rangle+|0,e,e\rangle)/\sqrt{2}$ (see Fig.~\ref{fig5}).
Moreover, if we can prepare two qubits in their excited states, a
single photon output jointly emitted by two qubits can also be
obtained via this method. Therefore, this proposal might also be
exploited for quantum information processing (including error
correction codes~\cite{Stassi2017}) and quantum optics in the
microwave regime.

\section*{acknowledgements}
We thank Anton Frisk Kockum and Salvatore Savasta for discussions
and useful comments. X.W. and H.R.L. were supported by the Natural
Science Foundation of China under Grant no. 11774284. A.M. and
F.N. acknowledge the support of a grant from the John Templeton
Foundation. F.N. was partially supported by the MURI Center for
Dynamic Magneto-Optics via the AFOSR Award No. FA9550-14-1-0040,
the Japan Society for the Promotion of Science (KAKENHI), the
IMPACT program of JST, JSPS-RFBR grant No 17-52-50023, CREST grant
No. JPMJCR1676, and RIKEN-AIST Challenge Research Fund.

%

\end{document}